\documentstyle[12pt,aasms4]{article}

\def\kms{km~s$^{-1}$}
\def\Msol{M$_\odot$}
\def\Lsol{L$_\odot$}

\def\lsim{\mathrel{\lower .85ex\hbox{\rlap{$\sim$}\raise
.95ex\hbox{$<$} }}}
\def\gsim{\mathrel{\lower .80ex\hbox{\rlap{$\sim$}\raise
.90ex\hbox{$>$} }}}

\def\vi{(V$-$I)}

\begin{document}

\def\thefootnote{\fnsymbol{footnote}}

\title{The Internal Kinematics of the Leo~I Dwarf Spheroidal Galaxy:
Dark Matter at the Fringe of the Milky Way
\footnote{Based on observations obtained at the W.M. Keck Observatory,
which is jointly operated by the California Institute of Technology
and the University of California.}}

\def\thefootnote{\arabic{footnote}}

\author{Mario Mateo\altaffilmark{1}}
\affil{\tt e-mail: mateo@astro.lsa.umich.edu}
\author{Edward W. Olszewski\altaffilmark{2}}
\affil{\tt e-mail: edo@as.arizona.edu}
\author{Steven S. Vogt\altaffilmark{3}}
\affil{\tt e-mail: vogt@ucolick.org}
\author{Michael J. Keane\altaffilmark{4}}
\affil{\tt e-mail: mk@ctio.noao.edu}

\altaffiltext{1}{Department of Astronomy, University of Michigan, 821
Dennison Bldg., Ann Arbor, MI \ \ 48109--1090}
\altaffiltext{2}{Steward Observatory, University of Arizona, Tucson,
AZ \ \ 85721} 
\altaffiltext{3}{Lick Observatory, University of
California, Santa Cruz, CA \ \ 95064} 
\altaffiltext{4}{Cerro Tololo Interamerican Observatory, Casilla 603, 
La Serena, Chile}

\begin{abstract}

We present radial velocities of 33 red giants in the Leo~I dSph galaxy
obtained from spectra taken with the HIRES echelle spectrograph on the
Keck~I telescope.  These data have a mean precision of 2.2 \kms\ and
lead to estimates of the central velocity dispersion and systemic
velocity of Leo~I of $8.8 \pm 1.3$ \kms, and $287.0 \pm 1.9$ \kms,
respectively.  The systemic velocity confirms past results that Leo~I
has an unusually large galactocentric velocity, implying the presence
of a massive dark halo in the Milky Way or an extended dark component
pervading the Local Group.  The V-band $(M/L)$ ratio of Leo~I is in
the range 3.5-5.6. We have produced a set of models that account for
the effects of stellar evolution on the global mass-to-light ratio of
a composite population.  Because Leo~I contains a dominant
intermediate-age population, we find that the V-band mass-to-light
ratio of Leo~I would be in the range 6-13 if it were composed
exclusively of old stars such as found in globular clusters.  This
suggests that Leo~I probably does contain a significant dark halo.
The mass of this halo is approximately $2 \times 10^7$ \Msol, similar
to the dark halo masses inferred for all other Galactic dSph galaxies.

Because Leo~I is isolated and has passed the Milky Way at most once in
the past, external tides could not plausibly have inflated its central
dispersion to the observed value.  We also considered whether MOdified
Newtonian Dynamics (MOND) could account for the internal kinematics of
Leo~I and conclude that this alternative gravitational model can
account for the Leo~I kinematics adequately without requiring a dark
halo.  The agreement with MOND is particularly good if the velocity
dispersion exhibits some anisotropy or the underlying stellar mass
function is only slightly different than a classical Salpeter law.

\end{abstract}

\clearpage

\section{Introduction}

Over the past few years, there has been steady progress in obtaining
reliable internal kinematic data for a number of the dwarf spheroidal
(dSph) satellites of the Milky Way (see recent reviews by Olszewski
1998; Mateo 1997, 1998).  One interesting result to emerge from these
studies is that the minimum mass of a dSph system is about $2.0 \pm
1.0 \times 10^7$ \Msol; consequently, the lowest-luminosity dwarfs are
also the ones with the largest $(M/L)$ ratios (Vogt et al 1995, Mateo
1997, 1998; Olszewski 1998).  We have also learned that only one dSph,
Ursa~Minor, is convincingly rotating with $v_{rot}/\sigma_0 \sim 0.5$
(Hargreaves et al. 1994; Armandroff et al. 1995).  Binary stars have
been detected among the red giants used for kinematic studies of dSph
galaxies (Mateo et al. 1991; Olszewski et al. 1995; Queloz al. 1995),
but their effect on the derived masses of these galaxies has been
shown to be negligible (Olszewski et al 1996; Hargreaves et al 1996).
At present, reliable estimates of the central velocity dispersions
have been obtained for all but one of the dSph satellites of the Milky
Way (Mateo 1998).  The lone exception -- and the subject of this paper
-- is the outermost system, Leo~I.

Because of its remote location in the outer halo of the Milky Way,
Leo~I is a particularly useful system to test whether tidal effects or
dark matter can best explain the kinematics.  Many models of the
dynamical history of the Local Group reveal that the birthplace of
Leo~I was far from any large galaxies of the Local Group (Byrd et al
1994; Peebles 1989, 1995).  It has also been suggested that Leo~I,
Leo~II and some other Milky Way dwarfs constitute a single `stream' of
galaxies -- mostly on the basis of their positions in the sky (Kunkel
1979; Lynden-Bell 1982; Majewski 1994).  However, when the kinematics
of the putative stream members are considered ($v_{helio, LeoI} = 287$
\kms; $v_{helio, LeoII} = 76$ \kms), the physical association is far
less compelling (Lynden-Bell and Lynden-Bell 1995).  Thus, Leo~I
appears to be a relatively isolated `test particle', one whose
internal kinematics are unlikely to have been significantly affected
by strong interactions with other systems.

Given our basic lack of understanding of the nature of the postulated
dark matter in galaxies, and the interesting successes of MOdified
Newtonian Dynamics (MOND) in explaining the properties of large
galaxies (most recently by MacGaugh and de~Blok 1998), it remains
important to test this concept in dwarf systems. Although MOND's
scaling parameter (Milgrom 1983a,b) was set to match rotation curves
of luminous galaxies, it has so far been consistent with the
observations of dwarf spheroidal galaxies (Gerhard 1994; Milgrom
1995).  Since MOND is falsifiable, one clearcut failure eliminates
this model from further consideration (Kuijken and Gilmore 1989; Lake
and Skillman 1989; Milgrom 1990).  Because the internal gravitational
accelerations within Leo~I are comparable to the limit where MOND is
postulated to be applicable, we can investigate here how well MOND
accounts for the internal kinematics of yet another low-mass dwarf
system.

This study represents the final chapter in the reconnaissance of the
kinematics of the known dwarf spheroidal companions of the Milky Way
(see Mateo 1998 for a review).  However, we are on the threshold of an
era of similar kinematic studies of more distant galaxies throughout
the Local Group and of far more extensive kinematic studies of our
closest neighbors.  In two subsequent papers, we shall explore the
kinematics of LGS~3 -- a companion of M~31 -- and carry out an
extensive study of the kinematics of Leo~I using fiber spectroscopy of
over 100 red giants in the galaxy (Cook et al 1998; Mateo et al. 1998,
hereafter Paper~III).

Throughout this paper we report all mass-to-light ratios in solar units.

\section{Observations and Reductions}

The data used in this study were obtained with the HIRES spectrograph
(Vogt et al. 1994) on the Keck~I 10m telescope during the nights of
March 13-14 (UT), 1996.  The observing and reduction procedures we
used are identical to those described in detail by Vogt et al (1995,
hereafter Paper~I).

Table 1 provides a log of our observations; a total of 40 spectra of
33 different Leo~I candidate red giants were obtained.  These
candidates were selected from a color-magnitude diagram of the central
regions of Leo~I obtained with the Michigan-Dartmouth-MIT Hiltner 2.4m
telescope in December 1992.  This diagram is reproduced in Figure~1
which also shows the selection criteria used to isolate candidate
Leo~I members.  Since Leo~I has a systemic velocity of about 286 \kms\
(Zaritsky et al 1989; Kulessa and Lynden-Bell 1992; Section 3.1) in a
direction where the Sun's reflex velocity due to Galactic rotation is
about 100 \kms, it is evident from Table~1 that every star we observed
is in fact a near-certain member of the galaxy.  Unlike the situation
for Fornax (Mateo et al. 1991), there is no kinematic ambiguity about
membership.  Nor are there any foreground red-dwarf contaminators as
in the case of Carina (Mateo et al 1993). Leo~I is located at a
relatively high Galactic latitude ($b = 49.1$) and its compact angular
size improves the contrast of members with respect to field stars.
The locations of the stars we observed are shown in Figure~2.
Figure~3 illustrates the velocity distribution of all stars in our
radial velocity sample for Leo~I.

We used the MDM photometric observations to obtain calibrated Johnson
V-band and Kron-Cousins I-band photometry of all of the
spectroscopically observed stars.  These frames along with data
obtained with the Palomar 1.5m telescope in 1991 were used to
determine precise coordinates for each star (details of this process
will be described in Paper~III where we present fiber spectroscopy of
nearly 100 Leo~I members).  We then calculated the position angle and
angular distance of each star from the center of the galaxy (taken to
be $\alpha_{2000} = 10^h 08^m 27.93^s$, $\delta_{2000} = 12^\circ 18'
18.4''$).  These photometric and astrometric results are provided in
Table~2 for each star that we observed.

The velocities and $R$ values (as defined by Tonry and Davis 1979)
listed in Table 1 for each spectrum were obtained using the {\tt
fxcor} package in version 2.10 of IRAF.  This routine uses a Fourier
cross-correlation algorithm to estimate velocities relative to a
high-S/N template spectrum.  Our template was composed of high-quality
spectra of numerous bright radial-velocity standards shifted to a
common velocity system.  We have found in many past studies that the
formal errors returned by {\tt fxcor} are overly optimistic by factors
of 2-5.  To better estimate the errors, we used our repeat
observations of six of the Leo~I stars, three stars in LGS~3 (Cook et
al. 1998), and multiple observations of standard radial velocity stars
to calculate the proportionality constant, $A$, for the relation
\begin{equation} \delta_i = A/(1+R_i)\end{equation}  
via a least-squares minimization technique (see Paper~I for details).
In equation 1, $\delta_i$ and $R_i$ are the formal error from {\tt
fxcor} and the Tonry-Davis $R$ statistic for star $i$, respectively.
This process gave $A = 14.9$ \kms.  Our estimates for the mean
velocities and velocity errors for the 33 stars we observed in Leo~I
are listed in Table~2.  The final mean velocities were determined
by weighting the individual measurements by $\delta^{-2}_i$ for stars
with multiple observations.

\section{Results}

\subsection{The central velocity dispersion of Leo~I}

The kinematic data listed in Table 2 can be used to calculate the
velocity dispersion, $\sigma_v$, of Leo~I.  The basic procedure has
been described in Paper~I and in reviews by Pryor (1992) and 
Mateo (1997).

The core radius of Leo~I is 3.3 arcmin (see Table~4).  Only one star
in our sample is located further from the center of Leo~I (star 1, $r
= 3.5$ arcmin) and the average angular separation of the stars we
observed from the center of the galaxy is 1.9 arcmin. Therefore, we
shall consider the velocity dispersion of our sample to be equivalent
to the central velocity dispersion of Leo~I; any corrections are
model-dependent and will be explored in detail in more comprehensive
models in Paper~III.

We have used here three model-independent estimators to calculate
$\sigma_v$.  The first is the weighted standard
deviation as defined by Mateo et al (1991, 1993) and in Paper~I:
$\sigma_{v,wm} = 8.6 \pm 1.2$ \kms. This value explicitly accounts for
the different velocity errors for each star.  The second estimator is
known as the bi-weight and is designed to be more robust than a
weighted standard deviation by minimizing the contribution of
outliers (Beers et al 1990).  For our observations, $\sigma_{v,bi} =
9.2 \pm 1.6$, where the error represents a 90\%\ confidence limit.
For this estimator, we gave each star equal weight.  Finally, a
maximum likelihood estimator was used to calculate the central
dispersion as described by Olszewski et al. (1996) and Armandroff et
al. (1995).  For this case, we derived $\sigma_{v,ml} = 8.8 \pm 1.3$.
It is clear that all three techniques provide essentially identical
estimates of $\sigma_{v}$ for our full sample.

We have searched for dependencies of the velocity dispersion for
different subsamples of stars. Table~3 lists $\sigma_{v}$ and systemic
velocities ($\langle v\rangle$) for (a) AGB stars: the 18 stars
brighter than I = 18.1, (b) RGB stars: 15 stars fainter than I = 18.1,
(c) the innermost third of the sample: the 11 stars closest to the
center of Leo~I, (d) the middle third of the sample: 11 stars, (e) the
outermost third of the sample: 11 stars, (f) the innermost half of the
sample: 16 stars, (g) the outermost half of the sample: 17 stars, (h)
the `eastern' sample: 16 stars, and (i) the `western' sample: 17
stars.  These eastern and western sample are defined with respect to
the least-squares rotation solution based on equation 2 and described
below in section 3.2.

To determine whether any of the differences in the velocity
dispersions of the various samples are statistically significant, we
also determined $\sigma_v$ and $\langle v\rangle$ for
randomly-selected samples of 11 and 16 stars drawn from the full
dataset.  In each case, we list the 5\%\ and 95\%\ values in these
distributions; for a normal distribution these percentile limits
approximately correspond to 2-sigma deviations from the mean.  These
random samples suggest that virtually none of the differences seen in
the dispersions for the subsamples listed in Table~3 are statistically
significant at greater than a 2-sigma level.  For example, the rather
large velocity dispersion of the outer-half sample is exceeded about
6\%\ of the time by a random sample of 16 stars.  Of course, the outer
and inner samples are strongly anti-correlated: If one subsample
exhibits a relatively large dispersion relative to the mean dispersion
for all stars, the other subsample must exhibit a small value.  We
conclude from Table~3 that there are no obvious variations in the
kinematic properties of Leo~I as a function of stellar type or
location.  The only possible exception is the bi-weight estimate of
the dispersion from the outer-third sample: $\sigma_v = 4.7 \pm 1.4$
\kms.  In this case, the bi-weight ignored (by design) two of the
stars with velocities furthest from the mean.  Since the other two
dispersion estimators obtained significantly larger values for
$\sigma_v$, we attach little significance to this one case.

Figure~4 is a plot of the observed radial velocities as a function of
stellar magnitude, color, radial distance, and position angle.  
There is no apparent trend visible in any of these plots, a conclusion
that is confirmed from least-squares fits of straight lines to these
data. In each case, the 2-sigma formal error of the best-fit slope
is larger than the slope itself.

Our adopted value for the mean central velocity dispersion of Leo~I is
based on the full sample of stars for which we derive $\sigma_{v,0} =
8.8 \pm 1.3$ \kms.  This value is the average of the three estimates
listed on the first line of Table~3, weighted by $\delta^{-2}$.  We
shall use this value in our subsequent discussion where we estimate
the central mass density and total mass of Leo~I.

\subsection{Rotation of Leo~I}

Paltoglou and Freeman (1987) suggested that the presence of
significant rotation in dSph galaxies could be used as a means of
testing whether these systems evolved from dIrr galaxies via
ram-pressure stripping, or perhaps by blow-out during energetic
star-formation episodes.  To date, the only dSph galaxy that exhibits
evidence of significant rotation is Ursa Minor (Hargreaves et
al. 1994; Armandroff et al. 1995; Mateo 1998). Since UMi is the
closest dSph, it is possible that tidal effects could cause streaming
motions that can mimic rotation (Piatek and Pryor 1995; Oh et
al. 1995; Johnston et al. 1995), although the apparent rotation axis
is not consistent with this explanation (Armandroff et al. 1995).
Tidal effects are negligible for Leo~I (see section 5.2) since it is
located far from the Milky Way.  

We used two algorithms to try to detect bulk rotation in Leo~I.  First, we
sorted the stars by position angle, then used each star to define an
axis passing through its position and the galaxy center.  We then
calculated the mean velocity and dispersion on either side of this
axis and calculated the mean velocity difference (and formal error of
the difference) of the two sides.  In no case did the velocity
difference exceed 0.7-sigma from which we conclude there is no
detectable rotation.  

The second approach was to perform a least-squares fit to a
cylindrical solid-body rotation curve:
\begin{equation} v_i = v_0 + S_v R_i \cos\theta_i, \end{equation} 
where $v_0$ is the systemic velocity, $S_v$ is the rotational velocity
gradient, $R_i$ is the radial distance of star $i$ from the galaxy
center, and $\theta$ is the angle between the position angle of a star
$i$ and the line that is normal to the rotation axis and passes
through the galaxy center.  In order to transform this into a linear
least-squares problem, we determined $S_v$ for different assumed
rotation axes in increments of 0.1 deg.  The best fit gave an inferred
rotation axis 144.1 deg (measured from north through east) and a
systemic velocity of 287.4 \kms.  However, for this fit $\chi^2_{min}
= 18.3$ (per degree of freedom), and the formal error of the rotation
gradient was 0.017 $\pm$ 0.0096 \kms~arcmin$^{-1}$.  The mean velocity
difference of stars on both sides of the best-fit rotation axis is
listed in Table~3 for the three velocity estimators; in no case is the
difference larger than 1.1-sigma.  These results suggest that the
detected `rotation' is statistically insignificant.  This is visually
confirmed in Figure~5 which shows the Leo~I radial velocities versus
the radial separation projected onto the axis of maximum gradient for
the best-fit rotation solution (which is also shown with its formal
error).  Also plotted in Figure~5 is the reduced $\chi^2$ as a
function of rotation position angle showing the small variation and
large mean value of this statistic.

%C. Pryor has also kindly provided his two programs designed to measure
%rotation in dwarf galaxies using these sorts of data.  Both of his
%programs use essentially the same basic methods described above but
%adopt different algorithms to determine the best-fitting rotation
%models, and to determine the significance of the fits; more details
%can be found in Armandroff et al (199x).  These programs also failed to
%detect significant rotation in Leo~I with these data.

These negative results do {\it not} rule out global rotation in Leo~I.
Since the galaxy exhibits a relatively large central velocity
dispersion and because our sample subtends a limited radial
distribution, mostly within one core radius of Leo~I, we would have
only been able to detect rotation in the presence of a rapidly rising
rotation curve of moderate amplitude.  A more definitive discussion of
the rotation of Leo~I requires a larger, more extended sample of
giants (Paper~III).

\subsection{The systemic velocity of Leo~I}

As with the velocity dispersion, we have used a weighted mean, the
biweight, and the maximum likelihood technique to estimate the
systemic velocity of Leo~I.  The results for the full sample are:
$\langle v \rangle_{wm} = 287.3 \pm 1.6$ \kms, $\langle v \rangle_{bi}
= 286.7 \pm 2.0$ \kms, $\langle v \rangle_{ml} = 287.0 \pm 1.6$ \kms.
The systemic velocity of Leo~I is listed for various other subsamples
in Table~3.  The rotation analysis in the previous section also
provides an estimate of the systemic velocity: $v_{sys,rot} = 287.4$
\kms.  We also explored the possibility that the systemic velocity
might prove to be a function of stellar subsample in our dataset;
Table~3 and Figure~3 illustrate that the systemic velocity of Leo~I
does not significantly depend on subsample.  In particular, no
subsample exhibits a systemic velocity outside the 90\%\ confidence
limits determined from our Monte-Carlo simulations.  For comparison,
Zaritsky et al. (1989) obtained a mean velocity of $285 \pm 2$ \kms\
from 10 observations of six stars, while Kulessa and Lynden-Bell
(1992) report a mean heliocentric velocity of 290 \kms\ but with
no details or error estimate.  We confirm
these earlier estimates of the systemic velocity of Leo~I.

Zaritsky et al. (1989) showed that a modified timing argument
including the Milky Way, M~31 and Leo I gave a result consistent with
the classical timing argument, namely, that the Milky Way mass must be
large. This model assumes that Leo I receded from its original
galactocentric distance from universal expansion, approached the Milky
Way because of the overdensity of the Local Group, and is now receding
on a bound orbit. Models in which Leo I is unbound to the Milky Way
but remains bound to the Local Group (Byrd et al. 1994) also require a
considerable dark matter component spread throughout the volume of the
group.  It seems clear that a satisfactory understanding of the large
systemic velocity of Leo~I requires a comprehensive model of the
entire Local Group, perhaps also including the past effects of
galaxies now located in external, nearby groups (e.g. Peebles 1989,
1995).

\section{The Mass and Mass-to-Light Ratio of Leo~I}

Because our new observational data consist exclusively of velocities
of stars located close to the center of Leo~I, we shall only consider
single-component dynamical models in this paper.  We will consider
single- and two-component models in Paper~III where we analyze less
precise velocity measurements of a larger sample of stars located well
beyond the core radius of the galaxy.

By `single-component' we mean that the mass distribution is assumed to
follow that of the visible component.  The latter is determined from
star counts or surface brightness measurements.  A unique complication
in determining the structure of Leo~I is its proximity in the sky to
the bright star Regulus (Hodge 1971).  The most recent study (Irwin
and Hatzidimitriou 1995; hereafter IH95) used deep star counts on
Palomar Schmidt plates; the structural parameters they derived are
listed in Table~4 along with additional parameters from other sources.

The standard analysis for a single-component model -- often referred
to as the `core-fitting' or King's method (King 1966) -- is described
by Richstone and Tremaine (1986) who further adopt an assumption of
isotropy in the velocity distribution.  Pryor and Kormendy (1990;
hereafter PK90) explore the effects of an anisotropic velocity
distribution on the final mass and $(M/L)$ ratio for single- and
two-component models where the visible and dark component, if any, can
have distinct spatial extents.  Though we shall deal with
single-component models here, we use the results from PK90 below to
set a lower limit on the central mass density of Leo~I
in extreme two-component models.

The basic analysis in this paper follows that of Paper~I where we 
used the following relations:
\begin{equation} \rho_0 = 166 \sigma_0^2 \eta / R_{1/2}^2, \end{equation} 
\begin{equation} M_{tot} = 167 R_c \beta v_s^2, \end{equation} 
\begin{equation} I_0 = S_0 / 2.1 R_{1/2}, \end{equation} 
which explicitly assume that mass follows light and that the velocity
distribution is isotropic (see King 1966; Richstone and Tremaine 1986;
Kormendy 1987; PK90 for details on the basis of these relations).  The
quantities $\rho_0$, $M_{tot}$, and $I_0$ are the central mass, total
mass, and luminosity density for the model, respectively.  $\sigma_0$
is the central velocity dispersion as defined in Table~3, while $v_s$
is the so-called `scale velocity' defined by PK90.  $S_0$ is the
central surface brightness, expressed in units of \Lsol pc$^{-2}$,
while $R_c$ and $R_{1/2}$ are the King core radius and half-surface
brightness radius, respectively.  The other factors and parameters
($\eta$, $\beta$, and the scaling factor relating $v_s$ and $\sigma_0$)
are taken from the appropriate King models that best fit the
star-count profile of Leo~I.  In all cases, the radii are geometric
means, defined as $R_{geom} = R_{maj} (1 - e)^{1/2}$, where $e$ is the
ellipticity of the galaxy, $e = 1 - R_{min}/R_{maj}$.  The
concentration of Leo~I is defined as $\log R_t/R_c$ = 0.58, where
$R_t$ is the tidal radius; the appropriate values of the various
parameters for this model are also listed in Table~4 for the isotropic
case.  Note that if we retain the assumption that mass follows light
but drop the assumption of isotropy, the final effect on the inferred
mass and central density is relatively slight ($\sim$50\%\ lower
values of $\rho_0$ and $M_{tot}$; Merritt 1988; PK90); however, in this
case there is no guarantee that the resulting models will be dynamically 
stable (Binney and Mamon 1982; Merritt 1988).

Table~5 lists the masses and $(M/L)$ ratios we derive for Leo~I using
King's method and our new estimate of the central velocity dispersion.

\section{Discussion}

\subsection{The Case for Dark Matter in Leo~I}

The results in Table~5 do not strongly support the notion that Leo~I
contains a dominant DM component. A mass-to-light ratio of 4-8 is
smaller than values observed in more extreme cases of DM in dSph
systems such as Draco, Ursa~Minor, Carina, or Sextans where $(M/L)$
ratios ranging from 50-100 are seen (see Mateo 1998 for a review).  And
although the $(M/L)$ ratio we find for Leo~I is larger than the mean
value observed in low-concentration globular clusters -- $\langle
(M/L)_{0,V} = 1.51 \pm 0.10$ -- the difference is only barely
significant at the 2-sigma level\footnote{The results for the globular
clusters are taken from the 36 clusters with concentration parameters
$\leq 2.0$ compiled by Pryor and Meylan (1993).  Internal dynamical
effects may significantly alter the core stellar populations, and
hence the $(M/L)$ ratios, in more concentrated clusters (Djorgovski et
al. 1991; Shara et al. 1998).  The mass-to-light ratios we have
adopted from the compilation by Pryor and Meylan (1993) were
determined for the globular clusters in their sample techniques
similar to the core-fitting method we have adopted for for Leo~I in
this study.  These authors also list `total' $(M/L)$ ratios for the
clusters in their sample but these were determined using a
considerably more elaborate model than the simple King method we use
in section 4.  Thus, we shall exclusively compare the former
globular-cluster $(M/L)$ ratios with our results for Leo~I and the
other nearby dSph galaxies.}.  Armandroff and Da~Costa (1986) explored
the range of $(M/L)$ ratios for pure stellar populations comprised
only of stars, their remnants, for a range of power-law initial mass
functions.  They concluded that such populations can produce V-band
$(M/L)$ values as high as 5-6, though such extreme values for $(M\L)$
required steep mass functions dominated by low-mass stars, or shallow
mass functions that produce many massive, but dark remnants from
high-mass stars.  For a Salpeter mass function, the upper limit on
$(M/L)_V$ is about 2.5, just consistent with the lowest value derived
here for Leo~I (Table~5).

There is an important complication in this discussion in the case of Leo~I.
Lee et al. (1993) and Gallart et al. (1998) have determined that Leo~I
contains a significant young population.  Because main sequence stars
obey a steep mass-luminosity relation ($L \propto m^3$) and because
stellar remnants have extremely high $(M/L)$ ratios, the global
$(M/L)$ ratio of a composite population increases with time (e.g.,
Elson et al. 1989).

How does the aging of a stellar population alter the global $(M/L)$
ratio of a galaxy with a complex star-formation history (SFH) such as
Leo~I?  To answer this, we have estimated mass-to-light ratios for a
composite stellar population formed from a reasonable stellar mass
function and for a SFH consistent with that observed for Leo~I
(Gallart et al. 1998).  The adopted SFH is shown schematically in the
first panel of Figure~6 (taken from Mateo 1998).  Our models involve
timesteps of 10 million years during which a sub-population of stars
is born with a mass spectrum of the form $dN(m) \propto m^{-x} dm$ and
upper and lower stellar mass limits, $m_u$ and $m_l$, respectively.
One set of models uses a single power-law mass function corresponding
to the Salpeter slope, $x = 2.35$, from 0.07 \Msol\ to 30 \Msol.  A
second, `composite' mass function takes the Salpeter slope for $m \geq
0.3$\Msol\, but then switches to $x = 0.0$ for 0.07\Msol $\leq m \leq$
0.30\Msol.  Stars more massive than 8 \Msol\ are assumed to evolve
into 1.4\Msol\ neutron stars, while stars less massive than 8 \Msol\
become 0.6 \Msol\ white dwarfs at the end of their lifetimes.  Both
sorts of remnants are assumed to have negligible luminosity.  The
main-sequence lifetimes of the stars are taken to be proportional to
$M^{-2.5}$ -- consistent with the mass-luminosity relation above --
and we assume the solar lifetime to be 10 Gyr.

We have tried to account for the contribution of red giants (which
have low $(M/L)$ ratios) using two parameters.  First, to constrain
the total number of giants we simply assumed that 0.5\%\ of the total
number of the most massive stars that have not yet turned into
remnants in a single star-formation episode will be seen as giants
today.  This procedure fails for very young populations (age $\lsim$ 1
Gyr) since the form and mean luminosity of the giant and red
supergiant branches evolve rapidly for smaller ages (Mermilliod 1981;
Ferraro et al. 1995).  In addition, the red giant luminosities are
constrained to be a factor, $\gamma_L$, times the luminosity of stars
at the main-sequence turnoff (MSTO) point.

Because the red giant region in the CMD does not evolve rapidly with
age for populations older than about 1 Gyr, $\gamma_L$ is likely to
vary with time as the MSTO luminosity steadily drops with increasing
age.  To calibrate this behavior, we have used the models of Elson et
al. (1989) who studied the evolution of the mass-to-light ratio of
single-aged populations.  For their oldest models (12-14 Gyr), Elson
et al. (1989) predicted V-band mass-to-light ratios ranging from 4-20
instead of 1-2 as observed in globular clusters.  In practice, we
found that if we let $\gamma_L$ vary linearly between 25 at 1 Gyr, and
75 for ages $\geq 10$ Gyr, our models produce a $(M/L)$-age relation
for single-age populations that closely matches the slope of the same
relation calculated by Elson et al. (1989) {\it and} achieve $(M/L)
\sim 1.5$ for old (14 Gyr) stars.  It would clearly be useful to
repeat this calculation with a complete set of modern evolutionary
models to self-consistently account for the role of giants in the
evolution of the $(M/L)$ ratio.  However, our approach allows us to
rapidly explore how a complex star-formation history affects the
evolution of a system's mass-to-light ratio.

Table~6 summarizes the results of our models where we have explored
the role of the mass function and star-formation history on the
$(M/L)$ ratio of three sample populations.  The first, labeled
`globular', is a system in which constant star formation occurred only
from 12-14 Gyr ago.  The second, labeled `young', is a population
where the stars formed at a constant rate from 1-2 Gyr ago.  Finally,
we include results for a SFH appropriate to Leo~I.  These three SFHs
are shown in the top row of Figure~6.  We also list in the final
column of Table~6 a parameter denoted as $\lambda$; this is the ratio
of the $(M/L)$ ratio for the globular cluster population divided by
the M/L ratio of the other populations for the same mass function.

We can see in Table~6 that $\lambda$ varies from about 1.75-2.31 for
Leo~I, depending on the form of the IMF.  This is the factor by which
the observed $(M/L)$ ratio for Leo~I must be multiplied to compare the
results for this galaxy with the $(M/L)$ ratio of a typical globular
cluster.  For the present discussion, we adopt the average of the
results in Table~6, $\lambda_{LeoI} = 2.03 \pm 0.3$.  This implies
that the V-band $(M/L)$ ratio of Leo~I would range from about 6-13 if
it were composed only of stars similar to those found in globular
clusters (see Table~5).  We conclude that Leo~I does in fact contain a
significant dark component.

We have calculated correction factors for the other dSph systems with
core kinematic measurements using the schematic star-formation
histories compiled by Mateo (1998); the adopted SFHs are plotted in
Figure~6.  These results are also listed in Table~6 for both the
Salpeter and composite mass functions.  Also listed in the table are
the approximate values of the central M/L ratios for these galaxies
corrected for population effects and converted to a scale on which the
$(M/L)_V$ ratio of the globular cluster population (as defined above)
is $\sim 1.5$.

In Figure~7a we show the V-band $(M/L)$ ratios vs the V-band
luminosities of the nine Galactic dSph galaxies and the three luminous
M~31 dSph systems with kinematic data (NGC~147, NGC~185, and NGC~205).
These results have not been corrected for evolutionary effects.  In
Figure~7b we plot the corrected $(M/L)$ ratios and corrected
luminosities for the Galactic dSph systems using the mean $\lambda$
values determined from Table~6; we applied no corrections to the M~31
satellites.  The galaxy luminosities have also been corrected by the
appropriate factor ($\lambda^{-1}$) prior to plotting them in
Figure~7b.  These corrected $(M/L)$ ratios 
correspond to the values these galaxies would have if their
entire stellar mass could be converted to a population appropriate for
a globular cluster and with the same stellar mass function.

Apart from Sgr (see Section 5.2), the data plotted in both panels of
Figure~7 are consistent with a simple model in which the galaxy
$(M/L)$ ratios can be fit to a relation of the form
\begin{equation} (M/L) = (M/L)_s + M_{DM}/L,\end{equation}
where $(M/L)_s$ is the intrinsic $(M/L)$ ratio of the stellar
population in the galaxy, $M_{DM}$ is the mass of the underlying dark
halo, and $L$ is the total V-band luminosity in solar units.  The
best-fitting relations are shown in Figure~7.  For both panels, we
find $M_{DM} \sim 2.0 \times 10^7$\Msol\ for $(M/L)_s = 1.5$.  The
function described in equation~7 is plotted in Figure~7 for these
values.  Leo~I fits this relation well in both plots.  Whatever
mechanism controls the location of a galaxy in the $(M/L)$-$M_V$ plane
seems to also be at work in Leo~I.  It appears that all dSph systems
contain dark matter, and that these dark halos are remarkably uniform.
We take this as further evidence that Leo~I has a dark component of
similar basic properties as those found in other dSph systems.

\subsection{The Case for Tidal Heating in Leo~I}

Kuhn and Miller (1989), Kuhn (1993), Kroupa (1997), and Klessen and
Kroupa (1998) have argued that tidal effects can induce resonances in
dSph systems that can artificially inflate their central velocity
dispersions, hence mimicking dark matter.  Pryor (1996) has argued
from a theoretical standpoint that such heating is difficult to
understand as an explanation for the large central velocity
dispersions observed in dSph galaxies.  Detailed $n$-body simulations
of encounters of dSph galaxies and the Milky Way at impact parameters
ranging from 10-50 kpc (Piatek and Pryor 1995; Oh et al. 1995;
Johnston et al. 1995) suggest that close encounters can produce
streaming motions in outer parts of the dwarfs.  These motions will be
seen as a systematic change of the mean velocity along the major axis
and can be incorrectly interpreted as rotation.  Despite this
streaming in the outer regions of the dwarfs, such close encounters
negligibly alter the central velocity dispersion.  Although strong
evidence for tidal extensions and perturbations in some of the nearby
dSph systems is accumulating rapidly (Mateo et al. 1996; Alard 1996;
Fahlman et al. 1996; Kuhn et al. 1996; Kleyna et al. 1998), there are
no convincing demonstrations yet of strong tidal heating of the core
of a dSph galaxy, apart from the Sgr system which appears to be in the
process of tidal disruption as it passes extremely close to the Milky
Way (Bellazzini et al. 1996; Mateo 1998; Mateo et al. 1998; though see
Ibata et al. 1997).

The central dispersion and inferred central $(M/L)$ ratio of Leo~I is
considerably higher than we would have expected in the absence of dark
matter or tidal heating.  However, the remote location and large
systemic velocity of Leo~I makes it difficult to understand how tides
could have had a role in this result.  Byrd et al.  (1994) have
specifically modeled the orbit of Leo~I within the Local Group over a
Hubble time.  They argue that the galaxy has passed only one large
system -- the Milky Way -- in its entire lifetime, about 2-4 Gyr ago.
This is consistent with the rather small tidal radius of Leo~I ($\sim$
1 kpc; see Table 4) which was presumably established at the galaxy's
closest approach to the Milky Way ($\sim$70 kpc according to Byrd et
al. 1994, though IH95 suggest that perigalacticon for Leo~I was about
20-30 kpc).  Models by Peebles (1989, 1995), though different in
detail, likewise suggest that Leo~I has had at most one past encounter
with a large galaxy of the Local Group. These results are completely
at odds with any resonance heating mechanism: In models by Kuhn and
Miller (1989) and Klessen and Kroupa (1998), significant core heating
does not occur until a galaxy has experienced several orbits around
the Galaxy on timescales $\lsim 1$ Gyr.

We conclude that tides have not had a significant role in heating the
core of Leo~I.  Their effect can be ruled out as the source of the
large inferred $(M/L)$ ratio of that galaxy.

\subsection{The Case for MOND in Leo~I}

The persistent lack of hard evidence regarding the true nature of dark
matter has become sufficiently irksome that astrophysicists have begun
to question even the most basic assumptions of our analysis in Section
5.1.  One example is known as MOdified Newtonian Dynamics (MOND) in
which Newton's law of gravity is modified to include a repulsive term
at low accelerations.  The reader interested in the surprising success
of this model to explain galaxy kinematics and some discussions on the
rather far-reaching implications of this idea is referred to
Bekenstein and Milgrom (1994), Felton (1984), Sanders (1997), and
McGaugh and de~Blok (1998); though see Lake and Skillman (1989) and
Kuijken and Gilmore (1989).  In this section, we simply consider if
MOND can account for the kinematics of Leo~I without resorting to DM.

Milgrom (1995) defined a parameter $\eta = (3\sigma_0^2/2
R_c)/(V_\infty^2/R)$, where $\sigma_0^2$ is the central velocity
dispersion of a dwarf galaxy, $R_c$ is its core radius, and $R$ is its
distance from the center of the Galaxy.  $V_\infty$ is the asymptotic
circular rotation speed at a distance $R$, taken here to be 220 \kms.
The parameter $\eta$ is thus a measure of the ratio of the internal
and external gravitational accelerations; in the case of Leo~I we
obtain $\eta = 2.5$ implying that this galaxy can be treated as an
isolated system.  The appropriate MOND relation to apply in this case is
\begin{equation} M_{tot,MOND} = 81 \sigma_0^4 / 4 a_0 G, \end{equation}
where $a_0 = 1.2 \times 10^{-8}$ cm s$^{-2}$ is the MOND acceleration
parameter derived from observations of rotation curves of giant disk
galaxies (Milgrom 1983a,b; Milgrom 1998), and $G$ is the gravitational
constant.  All terms in Equation~7 are in cgs units for the value of
$a_0$ noted above.  If we adopt $\sigma_0 = 8.8 \pm 1.3$ \kms\
(Table~4), then $M_{tot,MOND} = 7.6 \pm 4.5 \times 10^6$ \Msol.  For
the `total' V-band luminosity listed in Table~5, this corresponds to
$(M/L)_V = 1.6 \pm 1.1$ where the error reflects the uncertainty in
$\sigma_0$ and $L_V$.

In Section~5.1 we argued that the present-day V-band $(M/L)$ ratio of
Leo~I should be in the range 0.4-0.8 on a scale where a
globular-cluster population has $(M/L)_V \sim 1$-2.  The MOND M/L
ratio is very slightly higher than this.  One could interpret this to
mean that Leo~I requires a dark component even with MOND.  Like the
relations we used in Section~4 to apply King's method to Leo~I,
Equation~7 assumes an isotropic velocity distribution and that mass
follows light.  The second assumption follows immediately in MOND
since there is no dark component that might be distributed differently
from the luminous material.  However, an anisotropic velocity field
may modify the inferred MOND mass estimate. If the arguments of PK90
for models where mass follows light and strictly Newtonian gravity are
valid for MOND, then the total mass and central mass densities of an
anisotropic MOND model may be about 2 times smaller than for the
isotropic case (see section 6).  This implies $(M/L)_V \gsim 0.8 \pm
0.3$, consistent with the expected value of 0.5-0.8, and certainly so
if the underlying mass function is only slightly steeper or shallower
than the Salpeter law (see section 5.1).

We conclude that MOND plausibly accounts for the internal kinematics
of Leo~I without the need for any DM.  Lake \&\ Skillman (1989) argued
that MOND does not alleviate the need for DM in the rotating Local
Group dwarf irregular galaxy IC~1613 and Kuijken and Gilmore (1989)
suggest that the motions of stars perpendicular to the Galactic disk
are also at odds with MOND and direct observations of the minimum disk
surface mass density in the solar neighborhood.  However, we stress
that none of these cases represent fatal deviations from MOND
predictions.  Milgrom (1990) argued that
inclination and distance uncertainties for IC~1613 are sufficiently
large to remove the discrepancy with MOND.  In Leo~I the agreement
with the MOND prediction is sufficiently good that only a small degree of
anisotropy or a slightly modified mass function brings our observations
and the MOND prediction into remarkably good agreement.

\section{Summary and Conclusions}

We have obtained precise radial velocities of 33 giants in the remote
dSph galaxy Leo~I using the HIRES echelle spectrograph on the Keck~I
telescope.  These observations represent the final chapter of the
reconnaissance of the central kinematic properties of the Galaxy's
satellite dSph systems (Mateo 1998).

Using King's method, we derive a central V-band mass-to-light ratio of
$3.5 \pm 1.4$, or a `total' $(M/L)_V$ of $5.6 \pm 2.1$ (see Table~5).
These values are only slightly higher than the mean mass-to-light
ratio observed in Galactic globular clusters ($1.51 \pm 0.10$; Pryor
and Meylan 1993).  However, this simple comparison is not entirely
valid because Leo~I is dominated by a relatively young stellar
population.  To determine the magnitude of this effect, we generated a
set of simple models to account for stellar evolution on the $(M/L)$
ratio of a composite system.  On a scale where the globular-cluster
mass-to-light ratios are in the range 1-2, we find that the V-band
$(M/L)$ ratio of Leo~I is in the range 6-13.  Thus, we conclude that
Leo~I does appear to contain a significant dark component.

The properties of the dark halo inferred for Leo~I closely match those
found in other Galactic dSph systems (see Figure~7).  We find no
evidence for rotation in Leo~I, nor any significant dependence of the
kinematics on stellar type or location in the galaxy.  However, we
have relatively little leverage on rotation or spatial variations of
the kinematics of Leo~I because our rather small sample of 33 stars is
almost entirely located within one core radius of the center of Leo~I.
Our results confirm earlier findings that the minimum mass of dSph
systems appears to be about $2.0 \times 10^7$ \Msol.  Astronomers have
yet to discover any galaxy, anywhere in the local Universe with a
kinematically determined mass smaller than this (Lo et al. 1993 Young
and Lo 1996, 1997a,b; Mateo 1998).

We have considered whether tides or MOdified Newtonian Dynamics (MOND)
might account for the internal kinematics of Leo~I.  This galaxy is a
good test case for tides and MOND because it is well isolated from any
external gravitational effects from the Milky Way.  Because Leo~I has
had only one moderately close passage to any large galaxy (Byrd et
al. 1994), we conclude that tides or tidal resonances are unlikely to
have produced the kinematics we see in the galaxy.  We also find that
MOND can plausibly account for the internal kinematics of Leo~I
without the need for DM. This is especially true if there is a small
degree of velocity anisotropy in the galaxy, or if its underlying
stellar mass function is only slightly steeper or shallower than a
classical Salpeter law.

Much of our analysis implicitly assumes that the stellar velocities
are isotropically distributed in Leo~I and, in the dark-matter models,
that mass follows light.  PK90 explored the effects of relaxing these
assumptions for Ursa~Minor and Draco.  In particular, the lowest
possible central mass density in a system such as Leo~I corresponds to
a highly anisotripic case where the dark matter is much more extended
than the visible material (Merritt 1988).  Using the same assumptions
for Leo~I that PK90 used for UMi and Dra, this limit
corresponds to $\sim 0.08$ \Msol\ pc$^{-3}$.  For a stellar $(M/L)_V$
ratio in the range 0.5-0.8, this limit implies that about 20-50\%\ of
the central mass density in Leo~I is dark ($I_{0,V} = 0.097$ \Lsol
pc$^{-3}$).  Note that in order to achieve such a low {\it central}
mass density for a King-like light profile as observed in Leo~I, a
considerable amount of mass must be found in an extended halo.  Thus,
it appears that we cannot use anisotropy in the velocity field to
fully eliminate the need for dark matter at all radii.  We shall
explore this issue in greater detail in Paper~III where we analyze
individually less precise kinematic results for a larger sample of stars
spread over a large radial range in Leo~I.

We confirm the large systemic velocity of Leo~I reported by Zaritsky
et al. (1989) and independently by Kulessa and Lynden-Bell (1992).
Interestingly, regardless of the final verdict on the DM content
within Leo~I, this large radial velocity implies a considerable DM
halo about our Galaxy if the dwarf is bound to the Milky Way (Zaritsky
et al. 1989; Kochanek 1996), or a large overall DM content of the
Local Group if Leo~I is bound to the group as a whole rather than any
individual giant galaxy (Byrd et al. 1994).  It will be intriguing to
determine the relationship, if any, between Leo~I's DM halo and
the dark matter that pervades the halo of the Milky Way and the Local
Group as a whole.

We close by noting that Leo~I is the most distant system for which
individual stellar velocities of old red giants have been obtained
with a precision of $\lsim 2$ \kms.  With the advent of the new
generation of 8-10m telescopes, it has become possible to extend these
sorts of kinematic studies to the very limit of the Galactic halo, and
to start to consider probing further out into the Local Group.  Cook
et al. (1998) report on a first foray into this realm in their
kinematic study of the M~31 satellite LGS~3.  It will be exciting to
discover if the trends we have begun to see in Galactic satellites
hold as we push to more distant systems in different environments and
to compare stellar and gas kinematics in low-mass dwarf irregular
systems.  Perhaps most importantly, it shall be of considerable
interest to explore whether there really is a minimum dark-halo mass
of approximately $2 \times 10^7$ \Msol, and, if so, whether a large
population of essentially dark dwarf systems abound in the Local Group
and beyond.

\vskip2em

\section{Acknowledgements}

MM was partially supported by grants from the NSF during the course of
this research.  EO was partially supported by NSF grants AST-9223967
and AST-9619524.  We would all like to thank the excellent support by
the Keck mountain staff, especially Barb Schaeffer, and the UC Keck
TAC for providing time to carry out these observations.  We thank
C. Pryor for sending us his programs to detect rotation and which we
used to confirm our independent analysis.  We also thanks Denise
Hurley-Keller for helping produce Figure 2.  Finally, we are grateful
to Stacy McGaugh for his careful reading of the manuscript.

\clearpage

\centerline{\bf Figure Captions}
\vskip2em

\noindent{\bf Figure 1} -- A color-magnitude diagram of bright stars
in the central region of Leo~I.  The region from which we extracted
candidate Leo~I red giant members is enclosed by the dashed line; the
33 stars that we observed spectroscopically are shown as large solid
squares.

\vskip1em

\noindent{\bf Figure 2} -- A finding chart of the stars observed
spectroscopically in Leo~I.  The numbers refer to the star IDs listed
in Table~2. The field is approximately $4.9 \times 5.1$ arcmin along
the horizontal and vertical axes, respectively.  North is towards the
bottom, and east is towards the left.  Star 15 is the northern most
star in the compact grouping in which it is located, while star 17
is the southern star of a close pair.  The center of the field is 
located at $\alpha_{2000} = 10^h 08^m 27.93^s$, 
$\delta_{2000} = 12^\circ 18' 18.4''$.

\vskip1em

\noindent{\bf Figure 3} -- The distribution of velocities of the 33
stars observed spectroscopically.  There are no obvious outliers in
the sample.

\vskip1em

\noindent{\bf Figure 4} -- Plots of the observed heliocentric
radial velocities of the 33 Leo~I stars in our sample vs radial distance
from the galaxy center ({\it upper left}), position angle relative to the
galaxy center ({\it upper right}), the \vi\ colors of the stars ({\it lower
left}), and the apparent I-band magnitude of the stars ({\it lower right}).
There are no significant correlations of the mean velocity or velocity
dispersion with respect to any of these variables.

\vskip1em

\noindent{\bf Figure 5} -- {\it Top panel} \ A plot of the observed
heliocentric radial velocities vs the radial distance projected along
the axis with the strongest rotational signal as determined using the
least-squares procedure described in Section 3.2 (see the discussion
related to equation 2).  The slope of the velocity gradient is shown
along with the 2-sigma error bars.  The range enclosed by the error
bars includes a slope of zero, meaning that the data do not reveal any
significant evidence of rotation in the central region of Leo~I.  {\it
Lower panel} \ The variation of the reduced $\chi^2$ as a function of
position angle about the center of Leo~I for equation 2. This angle
measures the direction of the maximum velocity gradient due to
rotation.  Although there is a broad minimum centered at PA $\sim
54^\circ$, the large value of $\chi^2$ and the small range of
variation of this statistic over the full range of PA values is
another indication that we have not detected a significant rotation
signal in Leo~I.  For comparison, the major axis position angle is
$79^\circ \pm 3^\circ$ (Table~4).

\vskip1em

\noindent{\bf Figure 6} -- Plots of the adopted star-formation
histories (SFH) of a selected set of objects.  The upper row shows the
SFHs of Leo~I, a hypothetical globular cluster, and a `young'
population (see details in Section 5.1 for details).  Rows 2-3 show
the SFHs of the other eight Galactic dSph satellites.  In all cases,
only relative star-formation rates are plotted and time is shown in
units of Gyr where we have assumed the oldest possible population has
an age of 14 Gyr.  These SFHs are used to calculate the $(M/L)$
correction factors described in section 5.1 and listed in Tables~6 and
7.  The SFHs adopted here for the galaxies are taken from Mateo (1998).

\vskip1em

\noindent {\bf Figure 7} -- {\it Top panel} \ The variation of the
`total' V-band mass-to-light ratio of Galactic and M~31 dSph systems
as a function of visual absolute magnitude.  The data for the other
galaxies are taken from Mateo (1998) and Kleyna et al. (1998).  {\it
Lower panel} \ The same plot but now where the $(M/L)$ ratios and
absolute magnitudes have been corrected for stellar evolutionary
effects as described in section 5.1 and using the star-formation
histories plotted in Figure~6.  We have applied no corrections in this
panel for the three M~31 satellites: NGC~147, NGC~185, and NGC~205.
In both plots the dashed lines correspond to equation 7 with $(M/L)_s
= 1.5$, and $M_0 = 2.0 \times 10^7$ \Msol.  We have adopted the
results of Ibata et al. (1997) for Sgr; their analysis assumes that
Sgr is in dynamical equilibrium despite some evidence to the contrary
(Bellazzini et al. 1996; Mateo 1998).  The points for Sgr should be
regarded as upper limits to the true $(M/L)$ ratio of that galaxy.

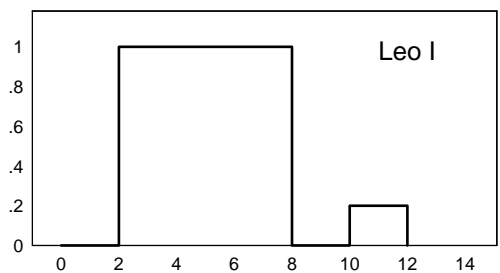
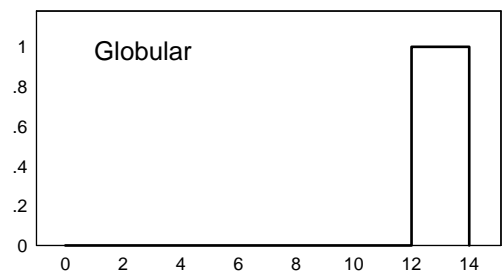
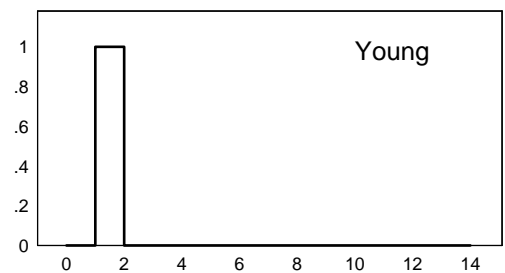
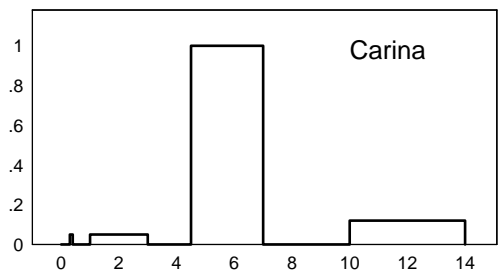
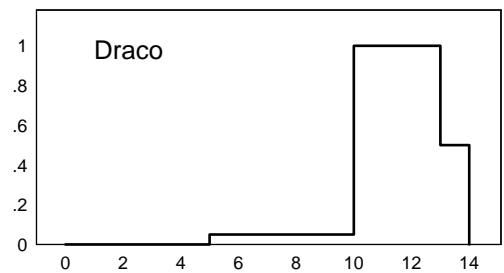
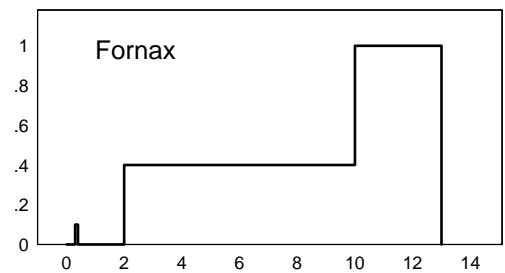
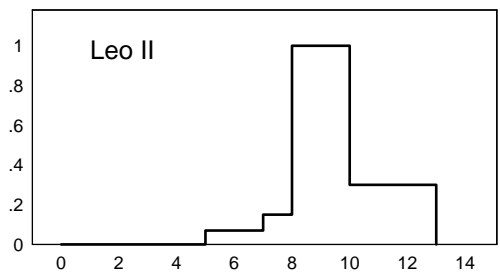
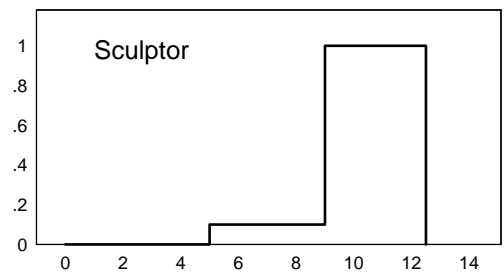
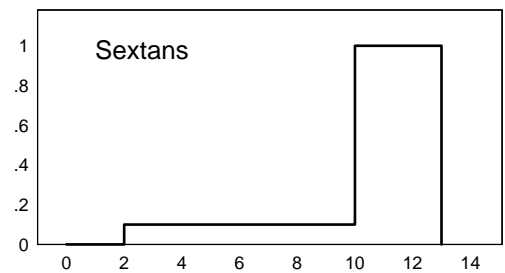
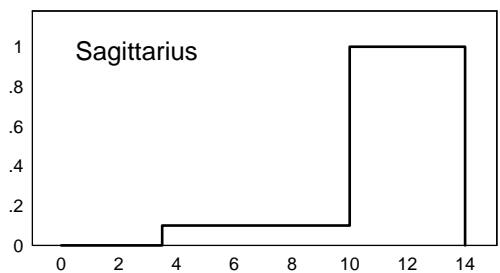
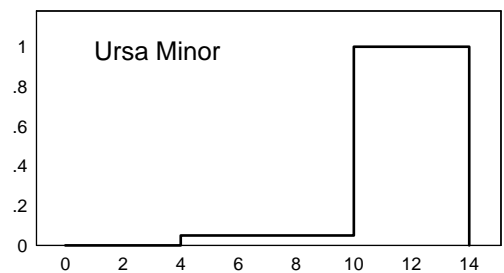
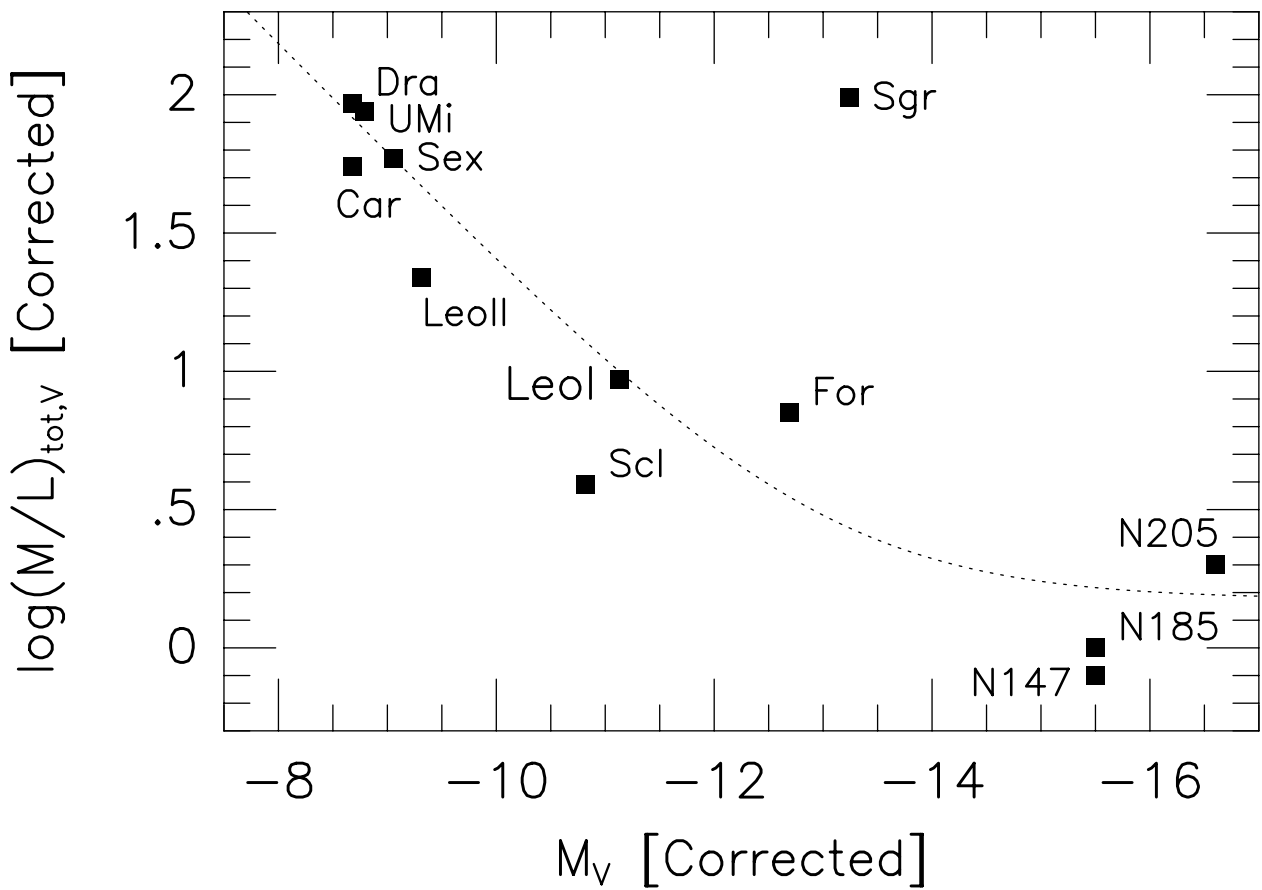
\end{document}